# Lifecycle of a sub-metered tertiary multi-use (GreEn-ER) building's open energy data: from resource mobilisation to data re-usability

Seun OSONUGA     Vincent IMARD     Benoit DELINCHANT     Frederic WURTZ

Univ. Grenoble Alpes, CNRS, Grenoble INP[1], G2Elab,

38000 Grenoble, France

March 15, 2024

## Abstract

The proliferation of sensors in buildings has given us access to more data than before. To shepherd this rise in data, many open data lifecycles have been proposed over the past decade. However, many of the proposed lifecycles do not reflect the necessary complexity in the built environment. In this paper, we present a new open data lifecycle model: Open Energy Data Lifecycle (OPENDAL). OPENDAL builds on the key themes in more popular lifecycles and looks to extend them by better accounting for the information flows between cycles and the interactions between stakeholders around the data. These elements are included in the lifecycle in a bid to increase the reuse of published datasets. In addition, we apply the lifecycle model to the datasets from the GreEn-ER building, a mixed-use education building in France. Different use cases of these datasets are highlighted and discussed as a way to incentivise the use of data by other individuals.

**Keywords**: Energy Data; Open Data Lifecycle; Data Reuse

---

[1] Institute of Engineering Univ. Grenoble Alpes

## Table des matières

# Introduction

In today's world, the importance of data cannot be overstated. Open data has been identified as one of the key enablers to the energy transition needed to address the rising global temperatures [1]. And now more than ever, even companies are struggling to deal with the enormous amounts of data that are generated in the processes [2]. As more data is being generated, it becomes important to ensure that data created or shared is fit for purpose, especially against the backdrop of limited resources. With the built environment accounting for almost 40% of greenhouse gas emissions, leveraging building data will be a key lever in the energy transition.

The turn of the decade has seen a large increase in the availability of building data, mostly in the residential domain [3]. Interesting projects such as COMBED [4] and the Building Data Genome Project [5] have striven to show the utility and difference between non-residential datasets and the more common residential datasets. However, these datasets often lack one key feature: there is not enough explanation of the lifecycle of these datasets and more machine-readable information about the individual buildings. Despite these shortfalls, these datasets have been used for many applications ranging from load forecasting to energy disaggregation. However, it is yet to be seen how the use of these datasets has modified the data collected and shared in many cases.

To this end, this article looks to propose the OPENDAL lifecycle, which builds on many of the fundamental concepts outlined in previous lifecycle models. It then goes ahead to adapt itself towards the concept of treating data to facilitate its reuse. The OPENDAL lifecycle details avenues to track data use as well as the feedback loops possible from the reuse of data by other parties. This lifecycle model is then used to characterise the movement of data from the GreEn-ER building, a sub-metered tertiary building in France used for education and research purposes.

# Definitions and State of the art in data lifecycles

Data Lifecycles have been described as the flow of data objects (files, links, or even databases) through different stages from their birth to their death [6]. These flows can be adapted to occur in a linear format or a cyclical manner [7].

One of the most prominent lifecycle models is that proposed by the US Geological Society [8], which linearly includes the steps: Plan, Acquire, Process, Analyse, Preserve, and Publish/Share. Although the proper deployment of resources in data creation is cited as one of the underlying challenges that lead to the lifecycle creation, it is not clear how the use of the data itself influences the creation of datasets. In contrast, the Digital Curation Center models proposed by [9] and [10] focus on cyclical lifecycles that take into account the circularity of data and the flow of information from the use to support new deployment. With the current push for open data, there has also been a recent proliferation of new data cycles targeting open government data and publishing with a purpose amongst many other things [11], [12], [13], [14]

However, it is worthy of note that despite the pivotal place that buildings play in the energy landscape today, there has not been a lifecycle that looks to explore the valorisation of building data cycles or energy data lifecycles in general. One of the peculiarities of building energy data is that unlike government data and weather data is based on resources that are essentially in the commons. Building data is typically the property of an individual or organisation. And although a case can be made for public buildings being in the commons, that is not the case for most buildings. Therefore, it seems pertinent to create a lifecycle that is built around data that is most often in the private property domain and does not lend itself immediately to being opened.

## Proposed Lifecycle

### Methodology for Development

To develop this data life cycle, the authors focused on the key elements from existing lifecycles such as: the USGSS cycle, Open Linked Government Data Cycle, and DCC data lifecycle [8], [9], [12]. However, the authors looked to simplify these for practitioners and clarify points where the flow of information between steps was not clear. This resulted in a lifecycle model that is similar enough to things in the state-of-the-art but easier to appropriate, especially in the context of predominantly private data.

### Description of the Open Energy Data Lifecycle (OPENDAL)

Figure 1 shows the current state of the conceived energy data lifecycle. It includes activities that are carried out by the data holder as well as other carried out by re-users and the potential interaction between them.

1. Mobilization: This step involves the gathering of human, physical, and financial resources necessary for the open energy data project. This is the step that is expressed as planning in many lifecycle models. However, this includes more than just deciding how to deploy these resources. It also includes alignment around the motivations for the data collection. This is the point where justifications are made for the resources to be used on the data project. It is important to note that often times, sharing and re-use of data are not the primary intents when deploying these resources. With this context, defining this step as mobilization instead of planning is more fit-for-purpose. Some non-technical aspects that are critical in this step are the things necessary to obtain the rights to the data. This could include filing appropriate data handling agreements and fulfilling privacy data handling requirements amongst others.

2. Data collection: After mobilization of the resources the next step is then the collection of the data. This involves all the physical processes involved in the capturing and transmission of the necessary data. In the case of quantitative data with meters, this is a relatively straightforward process. However not all energy data is quantitative and some of this necessary data is qualitative, being either survey data or even meta-data.

3. Data (pre)-processing: This involves all the data cleaning and exploration processes necessary to begin the value extraction process of the data. In the case of personal data, this also includes the aggregation, clustering, and (pseudo)-anonymisation of the dataset. cleaning, anonymisation, and obtaining rights. The outputs of this step can be useful data that the data holder uses for internal purposes and data that is shared publicly or privately with other parties.

4. Self-use: This involves using the dataset to achieve objectives that only serve the data holder. These could be running an energy management system or something else that does not meet the eye of the public. An important point of self-use here is that it uses private data that is not privy to the general public. This private data can also be used in the creation of use cases that will eventually be published which is discussed in the next step.

5. Open data sharing: This is the actual step for making the data available for other stakeholders. As this is an open data lifecycle, we will dwell mostly on this. Best practices on this step will involve sharing the data in a FAIR [Findable, Accessible, Interoperable, Reusable] manner [15]. This usually involves sharing the data in repositories or data catalogues.

6. Data referencing, incentivization, and self-use: This is a group of three steps that are useful for the same goal. The goal is the encouragement of other individuals or entities to use the data that has been shared. This includes publishing data papers, organising hackathons, and publishing the data holder's use-cases of the dataset. This can be seen to influence the use of the data rather than having an actual flow of information or resources into the external data cycle steps denote by the broken lines in Figure 1. The three individual steps are as flows:

    a. Use-case publishing: This connotes the sharing of ways that the data has been used. In the context where this data is available in the public domain, this will encourage other stakeholders to use the available data for their use cases. Research articles are a popular example of use-case publishing as scientists will be interested in the data used for a study if it is available.

    b. Data referencing: This typically involves writing a data paper or explanatory content (such as a website etc) on the dataset. This can also take the form of presentation of a dataset at conferences and other events.

    c. Data use incentivization: This include the proposition of activities or challenges using the data. This can take the form of datathons, challenges for communities such as Kaggle. The main difference between this and the referencing step being that the referencing looks to get "eyes" on the data while incentivization looks to get "hands" on the dataset.

7. Tracking use of data: This step involves the collection of information on the accessing and use of the shared dataset. This is an indispensable step to verify the actual reuse of the dataset but also ascertain its impact. This is a step that is very similar to the data provenance spoken about in data management. However, data provenance usually looks to describe the history of a dataset. This looks to describe what has come of the dataset. This is an essential step for the next step

8. Feedback from data use: This step involves reporting the impact of the datasets (use, relevance, etc.) to better design future mobilisation efforts. This however happens at two levels, internally and externally.

    a. Internal Feedback: This is usually generated from the internal use of the data by the data holder themselves. This usually addresses issues around data sufficieny, data formats, adequacy of processing step and other insights that the data holder can glean on the road to sharing the dataset with the public.

    b. External Feedback: This is in our opinion the most important step. This address all the same things that internal feedback does, however, by also incorporating the information from actual view and use by other stakeholders, it becomes possible to assess how

valuable the dataset is. This is a necessity that was already identified in previous work such as [16], [17]. And with this information of the value, the data holder can then make a good judgement or case for the mobilisation of resources. This can also be the case for the stopping of the allocation of resources and potentially moving the dataset to its end of life and archiving.

In addition to these core steps critical to the open data cycle, there are the things that happen with the data outside the control of the data holder. These form the environment of the data cycle and it is necessary to show clearly the links between this environment without including them exclusively. This renders the lifecycle more practicable.

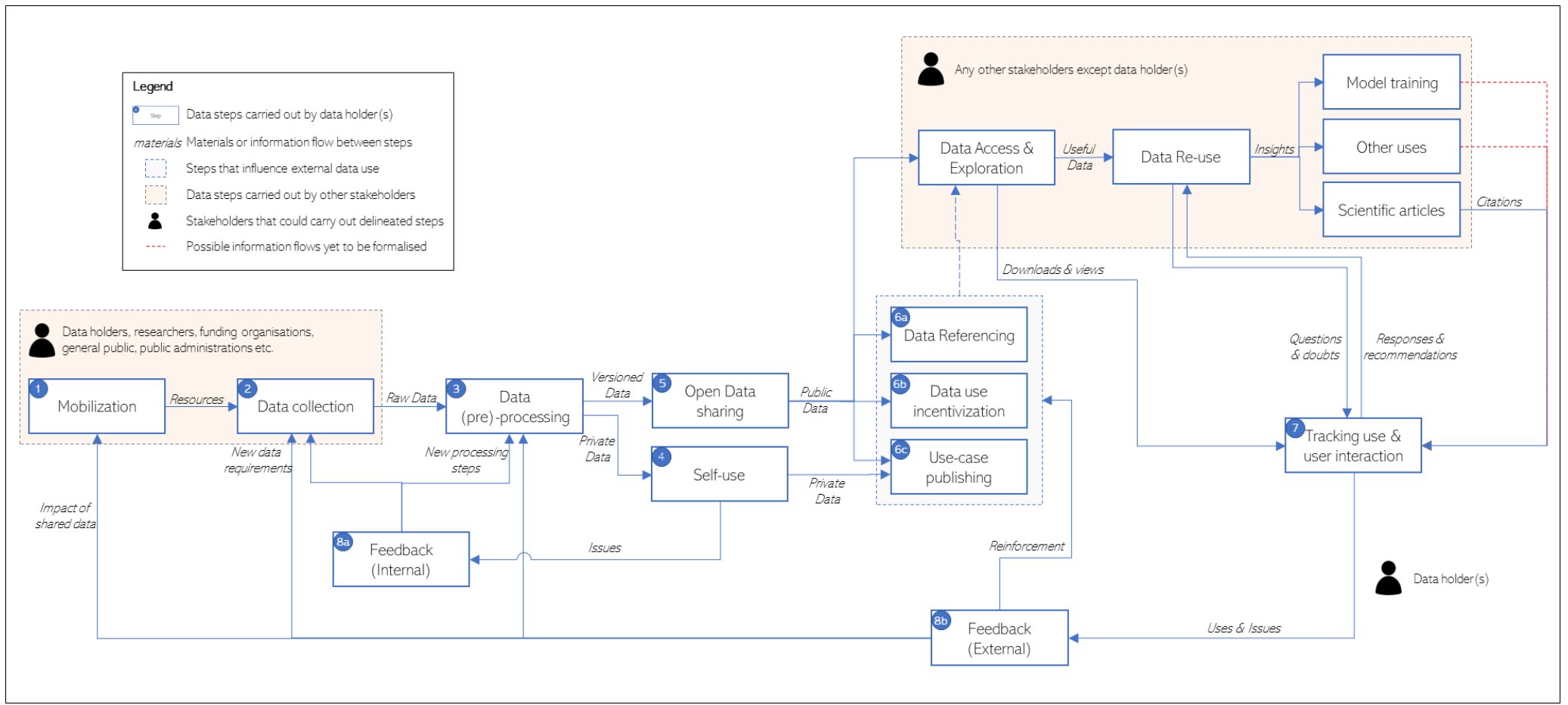

*Figure 1: Open Energy Data Lifecycle Schema.*

## Application of the lifecycle to data from the GreEn-ER Building

The GreEn-ER building is a multi-use tertiary building used for teaching and research that was completed in 2015 [18]. In addition to its classrooms, auditoriums, offices, and laboratories, it also features a restaurant and a library.

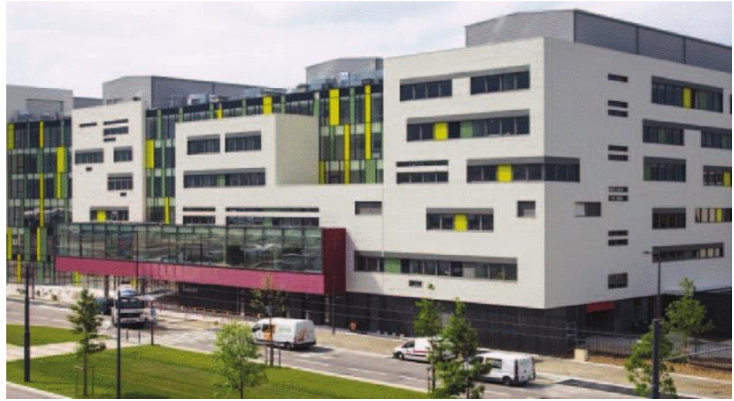

*Figure 2: GreEn-ER building, Grenoble*

### Mobilization

The GreEn-ER building was conceived as low-energy building to teach and show how energy should be managed. There was also the intention of setting up a living lab where different energy control and management strategies can be tested out [18]. The fruition of the project involved a public-private partnership which leveraged Eiffage for the daily energy management through a building performance agreement [19]. As a result, the building is equipped with over 2000 energy meters ranging from temperature sensors to occupancy sensors.

### Data Collection

The collection of the data from the building is managed by an SQL server integrated with the building management system supplied by Schneider Electric [20]. The desired logs are specified and kept in the database. In addition to data flows from real sensors, some datasets are created from the triangulation (by using formulas and models) of some datasets. These are referred to internal as "virtual meters".

### Data pre-processing

The pre-processing steps are determined by the end use required of the data. For example, in cases where we would be sharing the data for other researchers to work, there is barely a need to resample the data as it is shared in the most original way possible. However, to feed the many displays of energy use available in the GreEn-ER building the data has to be resampled and housed in a different database, Influx DB to allow effective real-time visualization as is done on the PREDIS-MHI website.

The legal requirements to share data from the GreEn-ER building are already fulfilled in the design of the building. One of these included the creation of a collection on recherche.data.gouv for the Grenoble Electrical Engineering Laboratory (G2Elab), the resident laboratory in the building.

The building datasets are being shared as results of the living lab experiment run at the G2Elab. These datasets will also be published in the collection of the OTE[2].

### Self Use

The data from the GreEn-ER building is used in two principal ways. The first is by Eiffage to ensure the proper functioning of the building. This include conducting analysis to ensure that the building energy system works as it is supported to. In addition to this, the data from the building is also displayed in the current areas to enlighten the building users of its impact. The second principal use of data from the building is for research purposes. In this regard, researchers in the G2Elab, laboratory on site, can use it for their various studies. These use cases will be detailed in the use case sharing sub-section below.

### Data Sharing

With the design of the GreEn-ER building as a living lab, sharing of data from the building has been on the main goal. At the moment, given the size and complexity of the different systems, the various datasets are been shared individually. There has the publishing of the electrical energy data and relevant metadata [21], thermal data from the PRESDIS-MHI platform (a sub-metered net-zero section of the building) [22], and the meteorological data from the weather station on the building[23]. The metrological data from the building is easily generalizable for the Presqu'île district, which was conceived as an eco-district.

In addition to sharing the data from the building through open channels, an separate section was created with SG-Interop place to share data with NTNU as part of the living lab concept [24]. This shared space allows researchers at NTNU to send control signals to the HVAC system and test different optimization and prediction models.

### Referencing/Incentivisation/Use case sharing

The datasets from the GreEn-ER building have been given light to by mostly the publication of use-cases which falls in the scope of our work as a research lab. The leading use cases have been around data processing and prediction algorithms. Some of these use cases involve addressing data quality issues [25], edge computing techniques [26], indoor temperature prediction [27], PV energy forecasting [28], and the comparison of multiple prediction models [29]. Some of the other paths of research covered include non-intrusive load monitoring [30], [31], energy simulator calibration, energy flexibility studies [32] and potential services possible with instrumented buildings [33].

In addition to the publishing of use-cases in the form of research papers, the datasets from the GreEn-ER building has been referenced data papers [34]. These datasets also feature in the many courses taught in the resident engineering school on energy management and artificial intelligence.

### Tracking use and User Interaction

The tracking and quantification of use of the datasets is carried out using downloads, citations, and requests. As the GreEn-ER datasets have only recently been shared, there are not a lot of downloads or citations from outside the data holders. As the datasets mature and with this wave of sharing we expect uses to increase. However, there have growing requests for access to the data from readers of the different scientific use cases published.

---

[2] OTE – Observatoire de la Transition Energetique is an open science project at the University of Grenoble-Alpes. More info available at https://ote.univ-grenoble-alpes.fr/

### Feedback

Feedback on the process happens on two scales: internal feedback based on the use by the owners/holders of the data themselves and external feedback which is usually as a result of the download and use by other people. At the moment, most of the feedback comes from the use of the datasets internally by Eiffage for their energy management service or researchers in the course of their analysis. Some of the biggest takeaways have been modification in the data-processing steps such resampling methods and the verification of the proper functioning of the many meters used in the building.

## Conclusions

This article has proposed a new open energy data lifecycle (OPENDAL) which piggybacks on the previous lifecycles in literature. However, it has several aspects that are develops explicitly. These include the tracking of use, the description of activities that can be used to incentivise use of the data and both internal and external feedback loops. In addition to this, the lifecycle was used to describe the process for the available energy datasets from the GreEn-ER building, a mixed-use education building in France.

As future work, we intend to publish the rest of the datasets from the GreEn-ER building and actively track their use and incorporate the necessary feedback. In addition to this, this lifecycle will be looked to be applied to other datasets and data projects. Some interesting prospective datasets/projects include energy communities, living labs, and personal mobility datasets. The application to these datasets will further validate the appropriateness of the data life cycle to the primarily private datasets of public interest.

# References

[1] World Economic Forum, 'Harnessing AI to accelerate the Energy Transition', Geneva, White Paper, Sep. 2021. Accessed: Jul. 19, 2023. [Online]. Available: https://www3.weforum.org/docs/WEF_Harnessing_AI_to_accelerate_the_Energy_Transition_2021.pdf

[2] M. E. Arass, I. Tikito, and N. Souissi, 'Data lifecycles analysis: Towards intelligent cycle', in *2017 Intelligent Systems and Computer Vision (ISCV)*, Apr. 2017, pp. 1–8. doi: 10.1109/ISACV.2017.8054938.

[3] H. Kazmi, Í. Munné-Collado, F. Mehmood, T. A. Syed, and J. Driesen, 'Towards data-driven energy communities: A review of open-source datasets, models and tools', *Renewable and Sustainable Energy Reviews*, vol. 148, p. 111290, Sep. 2021, doi: 10.1016/j.rser.2021.111290.

[4] N. Batra, O. Parson, M. Berges, A. Singh, and A. Rogers, 'A comparison of non-intrusive load monitoring methods for commercial and residential buildings'. arXiv, Aug. 27, 2014. doi: 10.48550/arXiv.1408.6595.

[5] C. Miller and F. Meggers, 'The Building Data Genome Project: An open, public data set from non-residential building electrical meters', *Energy Procedia*, vol. 122, pp. 439–444, Sep. 2017, doi: 10.1016/j.egypro.2017.07.400.

[6] B. Plale and I. Kouper, 'The Centrality of Data', in *Data Analytics for Intelligent Transportation Systems*, Elsevier, 2017, pp. 91–111. doi: 10.1016/B978-0-12-809715-1.00004-3.

[7] C. L. Borgman, 'The Lives and After Lives of Data', *Harvard Data Science Review*, Jun. 2019, doi: 10.1162/99608f92.9a36bdb6.

[8] J. L. Faundeen *et al.*, 'The United States Geological Survey Science Data Lifecycle Model', *Open-File Report*, 2014, doi: 10.3133/ofr20131265.

[9] S. Higgins, 'The DCC Curation Lifecycle Model', vol. 3, no. 1, 2008.

[10] P. Constantopoulos *et al.*, 'DCC&U: An Extended Digital Curation Lifecycle Model', *IJDC*, vol. 4, no. 1, pp. 34–45, Jun. 2009, doi: 10.2218/ijdc.v4i1.76.

[11] S. I. H. Shah, V. Peristeras, and I. Magnisalis, 'DaLiF: a data lifecycle framework for data-driven governments', *Journal of Big Data*, vol. 8, no. 1, p. 89, Jun. 2021, doi: 10.1186/s40537-021-00481-3.

[12] L. Ding, V. Peristeras, and M. Hausenblas, 'Linked Open Government Data [Guest editors' introduction]', *IEEE Intell. Syst.*, vol. 27, no. 3, pp. 11–15, May 2012, doi: 10.1109/MIS.2012.56.

[13] Michiel De Keyzer, Nikolaos Loutas, and Stijn Goedertier, 'Training Module 2.1: The Linked Open Government Data & Metadata Lifecycle'. 2014. Accessed: Dec. 15, 2023. [Online]. Available: https://joinup.ec.europa.eu/community/ods/document/tm21-linked-open-government-data-metadata-lifecycle-en

[14] Andrew Young, Andrew J. Zahuranec, Stefaan G. Verhulst, and Kateryna Gazaryan, 'The Third Wave of Open Data Toolkit', The GovLab, New York NY USA, Mar. 2021. [Online]. Available: https://opendatapolicylab.org/third-wave-of-open-data/

[15] M. D. Wilkinson *et al.*, 'The FAIR Guiding Principles for scientific data management and stewardship', *Sci Data*, vol. 3, no. 1, Art. no. 1, Mar. 2016, doi: 10.1038/sdata.2016.18.

[16] F. Wurtz, B. Delinchant, L. Estrabaut, and F. Pourroy, 'Vers de nouvelles approches théoriques et pratiques pour la capitalisation des connaissances et la mise en réseau des compétences autour des modèles numériques pour le bâtiment : l'approche DIMOCODE', in *XXXe Rencontres AUGC-IBPSA*, Chambéry, France, Jun. 2012. Accessed: Feb. 15, 2024. [Online]. Available: https://hal.science/hal-00716983

[17] S. Hodencq, F. Forest, T. Carrano, B. Delinchant, and F. Wurtz, 'User Experience Inquiry to Specify COFFEE: A Collaborative Open Framework For Energy Engineering', in *ELECTRIMACS 2022*, S. Pierfederici and J.-P. Martin, Eds., in Lecture Notes in Electrical Engineering. Cham: Springer International Publishing, 2023, pp. 531–542. doi: 10.1007/978-3-031-24837-5_40.

[18] B. Delinchant, F. Wurtz, S. Ploix, J.-L. Schanen, and Y. Marechal, 'GreEn-ER living lab: A green building with energy aware occupants', in *2016 5th International Conference on Smart Cities and Green ICT Systems (SMARTGREENS)*, Apr. 2016, pp. 1–8. Accessed: Jan. 26, 2024. [Online]. Available: http://ieeexplore.ieee.org/abstract/document/7951366

[19] A. Lebas, 'Collaborations et partenariats public-privé: leviers de transition pour nos territoires?', 2022.

[20] B. Delinchant and J. Ferrari, 'Standards and Technologies from Building Sector, IoT, and Open-Source Trends', in *Towards Energy Smart Homes: Algorithms, Technologies, and Applications*, S. Ploix, M. Amayri, and N. Bouguila, Eds., Cham: Springer International Publishing, 2021, pp. 49–111. doi: 10.1007/978-3-030-76477-7_3.

[21] G. Martin Nascimento, B. Delinchant, F. Wurtz, P. Kuo-Peng, N. Jhoe Batistela, and T. Laranjeira, 'GreEn-ER - Electricity Consumption Data of a Tertiary Building', vol. 1, Sep. 2020, doi: 10.17632/h8mmnthn5w.1.

[22] S. Osonuga, S. Shahid, A. Chouman, F. Wurtz, and B. Delinchant, 'PREDIS-MHI Thermal Data'. Recherche Data Gouv, 2024. doi: 10.57745/TZDEIH.

[23] PREDIS-MHI (GreEn-ER), 'Météo GreEn-ER - GreEn-ER - Dashboards - Grafana', GreEn-ER Open Data Dashboard - Weather. Accessed: Mar. 06, 2024. [Online]. Available: https://mhi-srv.g2elab.grenoble-inp.fr/grafana/d/Ao6TxpsMz/meteo-green-er?orgId=3

[24] P. M. Papadopoulos *et al.*, 'Indoor thermal comfort analysis for developing energy-saving strategies in buildings', in *2023 International Conference on Future Energy Solutions (FES)*, Vaasa, Finland: IEEE, Jun. 2023, pp. 1–6. doi: 10.1109/FES57669.2023.10183297.

[25] J. O. C. P. Pinto *et al.*, 'Architecture Multi-Agents pour Surveiller la Qualité des Données de Consommation d'Énergie', presented at the Conférence IBPSA France, Reims, France, 2020. doi: HAL Id: hal-03345308.

[26] C. Cérin, K. Kimura, and M. Sow, 'Data stream clustering for low-cost machines', *Journal of Parallel and Distributed Computing*, vol. 166, pp. 57–70, Aug. 2022, doi: 10.1016/j.jpdc.2022.04.009.

[27] Z. Fang, N. Crimier, L. Scanu, A. Midelet, A. Alyafi, and B. Delinchant, 'Multi-zone indoor temperature prediction with LSTM-based sequence to sequence model', *Energy and Buildings*, vol. 245, p. 111053, Aug. 2021, doi: 10.1016/j.enbuild.2021.111053.

[28] B. Delinchant, V. Imard, Y. Bermudez, and F. Wurtz, 'Lightweight hybrid local PV short term forecasting using both physics and data', in *APPEEC, 15th Asia-Pacific Power and Energy Engineering Conference*, Chiang Mai, Thailand: IEEE PES, Dec. 2023. Accessed: Feb. 01, 2024. [Online]. Available: https://hal.science/hal-04339309

[29] V. H. Nguyen, Y. Besanger, and Q. T. Tran, 'Self-updating machine learning system for building load forecasting - method, implementation and case-study on COVID-19 impact', *Sustainable Energy, Grids and Networks*, vol. 32, p. 100873, Dec. 2022, doi: 10.1016/j.segan.2022.100873.

[30] G. F. Martin Nascimento, F. Wurtz, P. Kuo-Peng, B. Delinchant, and N. Jhoe Batistela, 'Quantifying Compressed Air Leakage through Non-Intrusive Load Monitoring Techniques in the Context of Energy Audits', *Energies*, vol. 15, no. 9, Art. no. 9, Jan. 2022, doi: 10.3390/en15093213.

[31] M. K. Akbar, M. Amayri, N. Bouguila, B. Delinchant, and F. Wurtz, 'Evaluation of Regression Models and Bayes-Ensemble Regressor Technique for Non-Intrusive Load Monitoring', *Sustainable Energy, Grids and Networks*, p. 101294, Feb. 2024, doi: 10.1016/j.segan.2024.101294.

[32] N. K. Twum-Duah, M. Amayri, S. Ploix, and F. Wurtz, 'A Comparison of Direct and Indirect Flexibilities on the Self-Consumption of an Office Building: The Case of Predis-MHI, a Smart Office Building', *Frontiers in Energy Research*, vol. 10, 2022, Accessed: Dec. 13, 2023. [Online]. Available: https://www.frontiersin.org/articles/10.3389/fenrg.2022.874041

[33] B. Delinchant, H. A. Dang, H. T. T. Vu, and D. Q. Nguyen, 'Massive arrival of low-cost and low-consuming sensors in buildings: towards new building energy services', *IOP Conf. Ser.: Earth Environ. Sci.*, vol. 307, no. 1, p. 012006, Jul. 2019, doi: 10.1088/1755-1315/307/1/012006.

[34] G. F. Martin Nascimento, F. Wurtz, P. Kuo-Peng, B. Delinchant, N. Batistela, and T. Laranjeira, 'GreEn-ER–Electricity consumption data of a tertiary building', *Frontiers in Sustainable Cities*, vol. 5, Aug. 2023, doi: 10.3389/frsc.2023.1043657.